\documentclass[superscriptaddress,showpacs,floatfix,twocolumn,prl]{revtex4}
\usepackage{graphicx}

\begin{document}
\begin{flushright}
LU TP 04-26\\
hep-lat/0406017\\
Revised October 2004
\end{flushright}
\vskip-2.4cm

\title{The Pseudoscalar Meson Mass to Two Loops \\ 
       in Three-Flavor Partially Quenched $\chi$PT
      } 

\author{Johan Bijnens}
\affiliation{Department of Theoretical Physics, Lund University,\\
S\"olvegatan 14A, S 223 62 Lund, Sweden}
\author{Niclas Danielsson}
\affiliation{Department of Theoretical Physics, Lund University,\\
S\"olvegatan 14A, S 223 62 Lund, Sweden}
\affiliation{Division of Mathematical Physics, LTH,\\
Lund University, Box 118, S 221 00 Lund, Sweden}
\author{Timo A. L\"ahde}
\affiliation{Department of Theoretical Physics, Lund University,\\
S\"olvegatan 14A, S 223 62 Lund, Sweden}


\begin{abstract}
This paper presents a first study of the pseudoscalar meson masses to 
two loops, or NNLO, 
within the supersymmetric formulation of partially quenched 
chiral perturbation theory 
(PQ$\chi$PT). The expression for the pseudoscalar meson mass in the case 
of three valence 
and three sea quarks with equal masses, but different from each other, 
is given to ${\cal O}(p^6)$, 
along with a numerical analysis.
\end{abstract}

\maketitle

\section{Introduction}

In the theory of the strong interaction, Quantum Chromodynamics (QCD),
it has so far been impossible to derive hadron masses
directly using analytical methods. Lattice QCD is an approach
whereby the functional integral is evaluated numerically to obtain various
physical quantities. One difficulty in this has been the inclusion of
dynamical quark effects, but this is an area where much recent progress
has been
made. However, for computational reasons, the masses of the valence
quarks can be
much more easily varied than those of the sea quarks.
This situation is referred to as partially quenched lattice QCD.

The quark masses which can be accessed by present simulations are,
especially for the sea quarks, significantly
larger than the physical up and down quark masses. 
In order to obtain the masses of the lightest hadrons, 
one thus needs to extrapolate from the
quark masses used in the lattice calculations to the physical ones.

In this domain Chiral Perturbation Theory ($\chi$PT), 
which is an effective low-energy
field theory approximation to QCD, should be valid. It provides the
correct quark mass dependence of the pseudoscalar meson masses, including
the nonanalytic dependences often referred to as chiral logarithms.
The three flavour $\chi$PT, as formulated by
Gasser and Leutwyler~\cite{GL1},
is valid in the QCD case with equal valence and
sea quark masses, and cannot be used to extrapolate from
partially quenched lattice QCD calculations. The extension
of $\chi$PT to the quenched~\cite{BG1} and partially quenched~\cite{BG2}
cases has been carried out by Bernard and Golterman. The most general
expression at one loop
for the pseudoscalar meson masses was worked out by
Sharpe and Shoresh~\cite{Sharpe1}. In this work we present the result for
the meson masses at next-to-next-to-leading order (NNLO) in PQ$\chi$PT.
It should be pointed out that the case of full QCD corresponds to the
special case of partially quenched QCD 
where the sea and valence quark masses are put equal.
The free parameters or low energy constants (LECs)
of $\chi$PT, i.e. the QCD case, may thus be unambiguously determined from 
those of PQ$\chi$PT. These points are discussed
in detail in \cite{Sharpe1}.

We work here in the version of PQ$\chi$PT without the supersinglet $\Phi_0$,
discussed in detail at one loop in~\cite{Sharpe2}. Here
we present the expression for the charged, or off-diagonal,
meson mass for the case of equal
valence and equal sea quark masses, but these different from each other.
The charged meson masses for the more general quark mass cases, 
as well as more details
of the present calculation will be presented elsewhere~\cite{BDL}.
Planned work includes the decay constants
and the neutral, or diagonal, meson masses.
Note that the presence of NNLO terms is already
seen in the works of ref.~\cite{FarchioniMILC}.

In the next sections we present the technical background, 
the full expression for the pseudoscalar meson mass, and finally a short 
discussion and numerical results
as a function of the input quark masses.

\section{Technical Overview}

Most of the technical aspects required for calculations at the two-loop
level in PQ$\chi$PT without $\Phi_0$ already exist, 
but they have not been mentioned as such. The full divergence structure 
at one loop was worked out in~\cite{CP1,CP2}.
There, it was noted that the divergence structure is really identical to 
that of normal $\chi$PT with the number of flavours equal to the number of
sea quark flavors, provided that the traces and matrices are replaced by the
supertraces and matrices appropriate for quenched $\chi$PT.
This is justified by the replica method. In ref.~\cite{DS} it
has been argued that the quenched approximation, i.e. no sea quarks, can 
also be obtained using the replica method. This entails a calculation with 
$n_F$ flavors, setting $n_F=0$ in the final answer.
In ref.~\cite{DS} this was proven explicitly at one-loop order. These arguments
can be generalized to higher orders and to PQ$\chi$PT.
This allows the use of the known results for $n_F$ flavours in normal 
$\chi$PT (valence and sea quark masses equal), in order to obtain the needed
Lagrangians at ${\cal O}(p^4)$ and ${\cal O}(p^6)$
from ref.~\cite{BCE1}, as well as the full divergence structure 
from ref.~\cite{BCE2}. In particular,
the $n_F$ flavor Lagrangian at ${\cal O}(p^4)$ contains the term with
$L_0$~\cite{BCE1}, see also~\cite{GL1}.
In the context of PQ$\chi$PT, this extra term was pointed out in 
ref.~\cite{Sharpe3}.
We will use the notation of refs.~\cite{BCE1,BCE2} for the Lagrangian at 
${\cal O}(p^4)$ and ${\cal O}(p^6)$,
$L_i^r,i=0,\ldots,10$ and $H_i^r,i=1,2$ for the renormalized 
${\cal O}(p^4)$ LECs, 
and $K_i^r,i=1,\ldots,115$ for the renormalized LECs at ${\cal O}(p^6)$.
The reason why these Lagrangians can be obtained
by simply replacing the trace by the relevant 
supertrace~\cite{BG1,BG2,Sharpe1} is that the structure of the equations
of motion
and all other identities used are the same as for the $n_F$ flavor case
of refs.~\cite{BCE1,BCE2}. However, the Cayley-Hamilton identity used there 
to reduce the Lagrangians for
the two and three flavour case is not valid in PQ$\chi$PT.

The last technical difficulty in PQ$\chi$PT as compared to the mass
calculations at two loops in $\chi$PT~\cite{ABT1,ABT2} are the extra two-loop
integrals needed since the propagators in PQ$\chi$PT can have a double pole
structure. These extra integrals can be evaluated using the same
methods as was used for the integrals in ref.~\cite{ABT1}, provided that 
derivatives of those expressions w.r.t. the relevant masses
in the propagators are taken. 

\section{The Pseudoscalar Meson Mass to ${\cal O}(p^6)$}

\begin{figure}[h!]
\begin{center}
\includegraphics[width=\columnwidth]{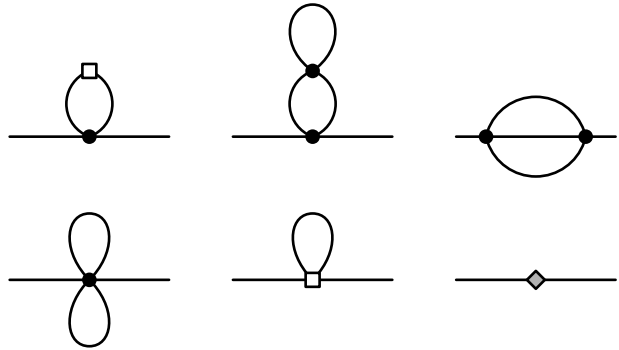}
\caption{Feynman diagrams at ${\mathcal O}(p^6)$ or two-loop
for the pseudoscalar
meson mass. Filled 
circles denote vertices of the ${\mathcal L}_2$ Lagrangian, whereas open
squares
and shaded diamonds 
denote vertices of the ${\mathcal L}_4$ and ${\mathcal L}_6$ Lagrangians,
respectively.}
\label{feynfig}
\end{center}
\end{figure}

The mass of the pseudoscalar meson is obtained by means of dimensional
regularization from the diagrams in Fig.~\ref{feynfig}
and those at ${\mathcal O}(p^2)$ and ${\mathcal O}(p^4)$.

The result is expressed in terms of
the valence 
quark mass, $m_{qV}$, and the sea quark mass, $m_{qS}$,
via $\chi_1 = 2 B_0 m_{qV}$, $\chi_4 = 2 B_0 m_{qS}$ and
$\chi_{14} = (\chi_1+\chi_4)/2$. These correspond to the lowest order
charged meson masses. Other parameters include the decay constant in the 
chiral limit ($F_0$), the quark condensate in the chiral limit, via 
$\langle{\bar q}q\rangle = - B_0 F_0^2$, and the LECs
of ${\cal O}(p^4)$ and ${\cal O}(p^6)$, 
the $L_i^r$ and $K_i^r$. 

The finite parts of the loop integrals in the expressions below
are
\begin{eqnarray}
\bar A(\chi) &=& -\pi_{16}\, \chi \log(\chi/\mu^2),
\nonumber\\
\bar B(\chi,\chi;0) &=& -\pi_{16} \left(1+\log(\chi/\mu^2)\right),
\nonumber\\
\bar C(\chi,\chi,\chi;0) &=& -\pi_{16}/(2 \chi)\, ,
\end{eqnarray}
where the subtraction scale dependence has been moved into the loop
integrals. We also define $\pi_{16} = 1/(16 \pi^2)$.
The finite two-loop sunset integrals $H^F,H_1^F,H_{21}^F$ that 
appear in the top right diagram of Fig.~\ref{feynfig} may be
evaluated using the methods of~\cite{ABT1}.
The notation used for the integrals is the same as in ref.~\cite{ABT1} 
except that an extra integer argument now indicates the
needed propagator structure. Index (1) corresponds to the case of
single propagators only, as in ref.~\cite{ABT1}, whereas index (2) 
indicates that the first propagator appears squared,
index (3) that the second propagator appears squared, 
and finally index (5) that the first and second propagators 
appear squared. Explicit expressions can be found in ref.~\cite{ABT1}, 
and by taking derivatives w.r.t. the masses of the expressions there.

We write the pseudoscalar meson mass
in the form
\begin{equation}
M_{\mathrm{PS}}^2 = \chi_1\,\left( 1 + \frac{\delta^{(4)}}{F_0^2} + 
\frac{\delta^{(6)}_{\mathrm{ct}} + \delta^{(6)}_{\mathrm{loops}}}{F_0^4} + 
\mathcal{O}(p^8) \right),
\end{equation}
where we have separated the ${\cal O}(p^4)$ and ${\cal O}(p^6)$ 
contributions.
The two-loop part $\delta^{(6)}$ has been further split into the
contributions from the chiral loops and from the ${\cal O}(p^6)$ counterterms.
The ${\cal O}(p^4)$ mass shift is
\begin{eqnarray}
\delta^{(4)} &=& 
        + \,24\,\chi_4\,(2\,L_6^r- L_4^r)
        + 8\,\chi_1\,( 2\,L_8^r-L_5^r)
\nonumber\\&&
        - 1/3\,\bar A(\chi_1)
        - 1/3\,\bar B(\chi_1,\chi_1;0)\,[\chi_1 - \chi_4],
        \label{d4tot}
\end{eqnarray}
which is in agreement with ref.~\cite{Sharpe1}. The ${\cal O}(p^6)$
mass shift due to the ${\mathcal L}_6$ Lagrangian is
\begin{eqnarray}
\delta^{(6)}_{\mathrm{ct}} &=& 
           - \,32\,\chi_1^2\,K_{17}^r 
           - 96\,\chi_1\chi_4\,K_{18}^r 
           - 16\,\chi_1^2\,K_{19}^r 
\nonumber\\&&
           - \,48\,\chi_1\chi_4\,K_{20}^r 
           - 48\,\chi_4^2\,K_{21}^r 
           - 144\,\chi_4^2\,K_{22}^r 
\nonumber \\&&
           - \,16\,\chi_1^2\, K_{23}^r
           + \,48\,\chi_1^2\,K_{25}^r 
           + 48\,[2\chi_1\chi_4 + \chi_4^2]\, K_{26}^r
\nonumber \\&&
           + \,432\,\chi_4^2\,K_{27}^r 
           + 32\,\chi_1^2\,K_{39}^r 
           + 96\,\chi_1\chi_4\,K_{40}^r. \label{d6ct}
\end{eqnarray}
If one further defines $\pi_{12}^2 = \pi^2/12 + 1/2$,
then the mass shift from the remaining ${\cal O}(p^6)$ terms is
\begin{widetext}
\begin{eqnarray}
\delta^{(6)}_{\mathrm{loops}}& = & 
     \quad\pi_{16}^2\:[15/32\,\chi_1\chi_4
        - 3/32\,\chi_1^2 + 73/64\,\chi_4^2] 
     - \pi_{16}^2\:\pi_{12}^2\:[41/9\,\chi_1\chi_4
       + 10/9\,\chi_1^2 + 17/4\,\chi_4^2] 
  \nonumber \\&&
   + \,\pi_{16}\:[26/3\,\chi_1\chi_4 - \chi_1^2 + 3\chi_4^2]\:L_0^r 
   + 4\:\pi_{16}\:\chi_1^2\:L_1^r 
   + \pi_{16}\:[2\chi_1^2 + 16\chi_4^2]\:L_2^r
   + \pi_{16}\:[17/3\,\chi_1\chi_4 - 5/2\,\chi_1^2 
   + 3/2\,\chi_4^2]\:L_3^r
  \nonumber \\&&
  + \,3\:\pi_{16}\:\bar A(\chi_{14})\:\chi_4 
  + 384\:\chi_1\chi_4\:L_4^rL_5^r - 1152\:\chi_4^2\:L_4^rL_6^r
  - 384\:\chi_1\chi_4\:L_4^rL_8^r + 576\:\chi_4^2\:L_4^{r2}
  \nonumber \\&&
  - \,384\:\chi_1\chi_4\:L_5^rL_6^r 
  - 128\:\chi_1^2\:L_5^rL_8^r + 64\:\chi_1^2\:L_5^{r2} 
  - \,8\:\bar A(\chi_1)\:[2\chi_1 - \chi_4]\:L_0^r 
  + 8\:\bar A(\chi_1)\:\chi_1\:L_1^r
  + 20\:\bar A(\chi_1)\:\chi_1\,L_2^r
  \nonumber \\&&
  - \,8\:\bar A(\chi_1)\:[2\chi_1 - \chi_4]\:L_3^r 
  + 16\:\bar A(\chi_1)\:\chi_4\:L_4^r
  + 16\:\bar A(\chi_1)\:[\chi_1 - 1/3\,\chi_4]\:L_5^r 
  + 16\:\bar A(\chi_1)\:[2\chi_1 - \chi_4]\:L_6^r 
  \nonumber \\&&
  + \,32\:\bar A(\chi_1)\:[\chi_1 - \chi_4]\:L_7^r 
  - 64/3\:\bar A(\chi_1)\:\chi_1\:L_8^r
  - \bar A(\chi_1)^2\:[13/72 - 5/18\,\chi_1^{-1}\chi_4]
  \nonumber \\&&
  + \,\bar A(\chi_1)\:\bar B(\chi_1,\chi_1;0)\:
      [77/36\,\chi_1 - 11/12\,\chi_4] 
  - 2/9\:\bar A(\chi_1)\:\bar C(\chi_1,\chi_1,\chi_1;0)\:
       [\chi_1\chi_4 - \chi_1^2] 
  + 24\:\bar A(\chi_{14})\:\chi_{14}\:L_0^r 
  \nonumber \\&&
  + \,60\:\bar A(\chi_{14})\:\chi_{14}\:L_3^r 
  - 48\:\bar A(\chi_{14})\:\chi_{14}\:L_5^r 
  + 96\:\bar A(\chi_{14})\,\chi_{14}\,L_8^r 
  - \bar A(\chi_{14})^2\:[27/4 - \chi_1\chi_{14}^{-1}] 
  \nonumber \\&&
  - \,2\:\bar A(\chi_{14})\:\bar B(\chi_1,\chi_1;0)\:\chi_4
  + 128\:\bar A(\chi_4)\:\chi_4\:L_1^r 
  + 32\:\bar A(\chi_4)\:\chi_4\:L_2^r 
  - 128\:\bar A(\chi_4)\:\chi_4\:L_4^r 
    \nonumber \\&&
  + \,128\:\bar A(\chi_4)\:\chi_4\:L_6^r 
  - \bar A(\chi_4)^2 
  + 8/9\:\bar A(\chi_4)\:\bar B(\chi_1,\chi_1;0)\:\chi_4
  + \bar B(\chi_1,\chi_1;0)\:[8\chi_1\chi_4 - 8\chi_1^2]\:L_0^r
    \nonumber \\&&
  + \,\bar B(\chi_1,\chi_1;0)\:[8\chi_1\chi_4 - 8\chi_1^2]\:L_3^r 
  + \bar B(\chi_1,\chi_1;0)\:[32\chi_1\chi_4 - 24\chi_4^2]\:L_4^r 
  - \bar B(\chi_1,\chi_1;0)\:[16\chi_1\chi_4 - 56/3\,\chi_1^2]\:L_5^r 
    \nonumber \\&&
  - \,\bar B(\chi_1,\chi_1;0)\:[48\chi_1\chi_4 - 32\chi_4^2]\:L_6^r 
  - \bar B(\chi_1,\chi_1;0)\:[32\chi_1\chi_4 - 16\chi_1^2
                             - 16\chi_4^2]\:L_7^r 
    \nonumber \\&&
  + \,\bar B(\chi_1,\chi_1;0)\:[64/3\,\chi_1\chi_4 - 32\chi_1^2 
                  + 16/3\,\chi_4^2]\:L_8^r 
  - \bar B(\chi_1,\chi_1;0)^2\:
         [1/3\,\chi_1\chi_4 - 5/18\,\chi_1^2 - 1/18\,\chi_4^2]
    \nonumber \\&&
  + \,\bar B(\chi_1,\chi_1;0)\:\bar C(\chi_1,\chi_1,\chi_1;0)\:[
    2/9\,\chi_1\chi_4^2 - 4/9\,\chi_1^2\chi_4 + 2/9\,\chi_1^3]
  - 16\:\bar C(\chi_1,\chi_1,\chi_1;0)\:
            [\chi_1\chi_4^2 - \chi_1^2\chi_4]\:L_4^r
    \nonumber \\&&
  - \,16/3\:\bar C(\chi_1,\chi_1,\chi_1;0)\:[\chi_1^2\chi_4 - \chi_1^3]\:L_5^r
  + 32\:\bar C(\chi_1,\chi_1,\chi_1;0)\:
          [\chi_1\chi_4^2 - \chi_1^2\chi_4]\:L_6^r
    \nonumber \\&&
  + \,32/3\:\bar C(\chi_1,\chi_1,\chi_1;0)\:
         [\chi_1^2\chi_4 - \chi_1^3]\:L_8^r
  + 5/9\:H^F(1,\chi_1,\chi_1,\chi_1;\chi_1)\:\chi_1
  - H^F(1,\chi_1,\chi_{14},\chi_{14};\chi_1)\:[\chi_1 - 1/4\,\chi_4]
    \nonumber \\&&
  + \,2\:H^F(1,\chi_{14},\chi_{14},\chi_4;\chi_1)\:\chi_4
  - 4/9\:H^F(2,\chi_1,\chi_1,\chi_1;\chi_1)\:[\chi_1\chi_4 - \chi_1^2]
    \nonumber \\&&
  - \,3/4\:H^F(2,\chi_1,\chi_{14},\chi_{14};\chi_1)\:
           [\chi_1\chi_4 - \chi_1^2]
  + 2/9\:H^F(5,\chi_1,\chi_1,\chi_1;\chi_1)\:[
    \chi_1\chi_4^2 - 2\chi_1^2\chi_4 + \chi_1^3]
    \nonumber \\&&
  + \,4\:H_1^F(3,\chi_{14},\chi_1,\chi_{14};\chi_1)\:
               [\chi_1\chi_4 - \chi_1^2]
  + 3/4\:H_{21}^F(1,\chi_1,\chi_{14},\chi_{14};\chi_1)\:\chi_1
    \nonumber \\&&
  + \,6\:H_{21}^F(1,\chi_4,\chi_{14},\chi_{14};\chi_1)\:\chi_1
  + 3/4\:H_{21}^F(2,\chi_1,\chi_{14},\chi_{14};\chi_1)\:[
    \chi_1\chi_4 - \chi_1^2].
         \label{d6loop}
\end{eqnarray}
\end{widetext}

As expected, 
the divergences in these expressions have cancelled. Note that in 
both eq.~(\ref{d4tot}) and eq.~(\ref{d6loop}), the extra pole structure
from the partially quenched nature disappears if we set $\chi_1=\chi_4$.
The remaining $\bar B(\chi_1,\chi_1;0)$ terms at ${\cal O}(p^6)$ in that
case are due to the fact that we have expressed the ${\cal O}(p^4)$
result in terms of the lowest order masses rather than the full physical 
masses, which was called unrenormalized in ref.~\cite{ABT1}. 
Consequently, they do not produce unphysical logarithms.

\section{Discussion and Conclusions}

\begin{figure}
\includegraphics[width=0.8\columnwidth]{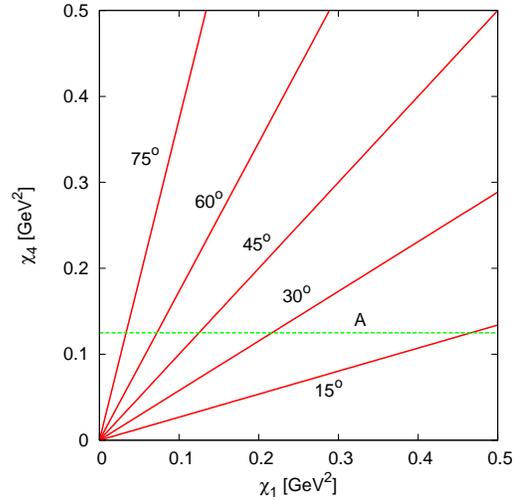}
\caption{\label{figlines} The lines in the $\chi_1$-$\chi_4$ plane
along which results in Figs.~\ref{figresult1}-\ref{figresult3} are shown.}
\end{figure}
In the long run, all input parameters should of course be determined by
a fit of the formulas of PQ$\chi$PT to lattice QCD data. At the present time,
we can only present results using input parameters from the continuum
work in $\chi$PT at the same order. We use as input parameters
the fit to data presented in ref.~\cite{ABT2}, called fit 10, which had
$F_0 = 87.7$~MeV and $\mu = 770$~MeV. Of the parameters
which were not determined there, we set $K_i^r=L_4^r=L_6^r=L_0^r=0$.
The last one
cannot be determined from experimental data and some recent results on
$L_4^r$ and $L_6^r$ can be found in ref.~\cite{piK}. 

\begin{figure}
\includegraphics[width=0.79\columnwidth]{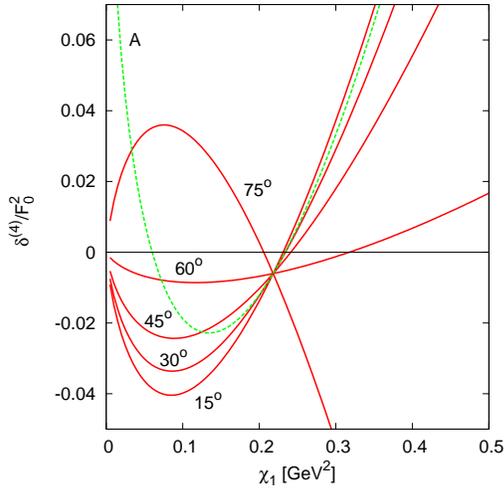}
\caption{\label{figresult1}
The result at ${\cal O}(p^4)$.}
\end{figure}
\begin{figure}
\includegraphics[width=0.79\columnwidth]{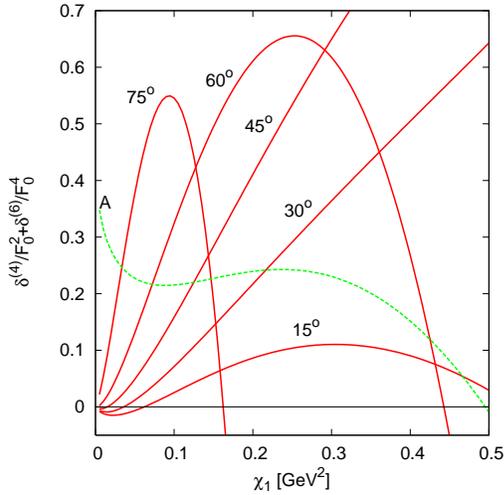}
\caption{\label{figresult2}
The sum of the ${\cal O}(p^4)$ and ${\cal O}(p^6)$ results.}
\end{figure}
\begin{figure}
\includegraphics[width=0.79\columnwidth]{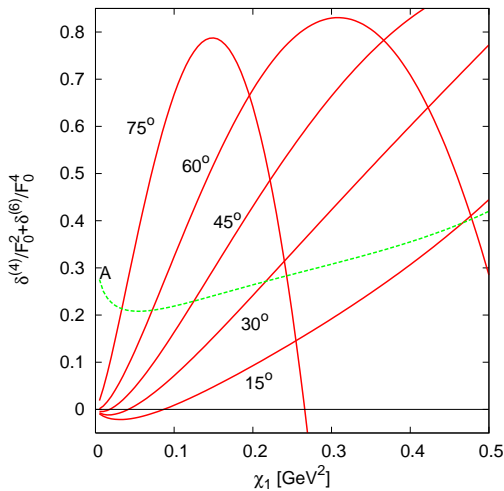}
\caption{\label{figresult3}
The sum of the ${\cal O}(p^4)$ and ${\cal O}(p^6)$ results
at $\mu=770~$MeV with all LECs set to zero.}
\end{figure}

In Fig.~\ref{figlines} we show the lines in the $\chi_1$-$\chi_4$
plane used in the following figures.
The remaining figures show the relative correction to the meson mass.
In Fig.~\ref{figresult1} we plot $M^2_{\mathrm{PS}}/\chi_1-1$ to 
${\cal O}(p^4)$, and in Fig.~\ref{figresult2} to ${\cal O}(p^6)$, in
both cases with the LECs set to the
fit 10 values of ref.~\cite{ABT2}. Fig.~\ref{figresult3} also shows the
result at ${\cal O}(p^6)$, but with the LECs set to zero.
Note that the effect of the unphysical logarithms is clearly
visible along the line labelled $A$ which has a constant sea quark mass.
That the ${\cal O}(p^6)$ contributions to the pseudoscalar meson masses
are sizable was expected.
Similar results for the case of $\chi$PT were obtained in 
refs.~\cite{ABT1,ABT2}, but cancellations with the contributions 
from the ${\cal O}(p^6)$ Lagrangian might change this. 

In conclusion, we have calculated the off-diagonal pseudoscalar meson mass
at NNLO order in PQ$\chi$PT, and have presented first results for the
case of equal valence quark masses and equal sea quark masses. 
The ${\cal O}(p^6)$ contributions are sizable, and the effects of the loop 
contributions are definitely nonnegligible at presently used quark masses 
in lattice QCD calculations.

\section*{Acknowledgements}

The program FORM 3.0 has been used extensively in these calculations
\cite{FORM}. This work is
supported in part by the Swedish Research Council
and the European Union TMR
network, Contract No. HPRN-CT-2002-00311  (EURIDICE).
TL acknowledges the Magnus Ehrnrooth foundation for a travel grant.

\end{document}